# Enhanced sources of acoustic power surrounding AR 11429

**Alina Donea, Christopher Hanson**

Monash Centre for Astrophysics, School of Mathematical Sciences, Monash University, Victoria 3800, Australia

E-mail: `alina.donea@monash.edu`

**Abstract.** Multi-frequency power maps of the local acoustic oscillations show acoustic enhancements ("acoustic-power halos") at high frequencies surrounding large active region. Computational seismic holography reveals a high-frequency "acoustic-emission halo", or "seismic glory" surrounding large active regions. In this study, we have applied computational seismic holography to map the seismic seismic source density surrounding AR 11429. Studies of HMI/SDO Doppler data, shows that the "acoustic halos" and the "seismic glories" are prominent at high frequencies 5–8 mHz. We investigate morphological properties of acoustic-power and acoustic emission halos around an active region to see if they are spatially correlated. Details about the local magnetic field from vectormagnetograms of AR 11429 are included.

We identify a $15''$ region of seismic deficit power (dark moat) shielding the white-light boundary of the active region. The size of the seismic moat is related to region of intermediate magnetic field strength. The acoustic moat is circled by the halo of enhanced seismic amplitude as well as enhanced seismic emission. Overall, the results suggest that features are related. However, if we narrow the frequency band to 5.5 – 6.5 mHz, we find that the seismic source density dominates over the local acoustic power, suggesting the existence of sources that emit more energy downward into the solar interior than upward toward the solar surface.

## 1. Introduction

Regions of enhanced high-frequency oscillation power are usually observed at the boundaries of active regions and are referred to as "acoustic halos". They show a local enhancement of oscillatory amplitude at the Sun's surface that could be due to enhanced local generation of acoustic noise or to a greater local sensitivity of the surface to acoustic noise arriving from distant sources. Properties of halos were studied by Braun et al. [1], Brown et al. [2], Hindman and Brown [3], Jain and Haber [4], Donea at al. [5, 6], more recently by Schunker and Braun [7] and Rajaguru et al. [8]. Khomenko and Collados [9] summarized the halos' properties including a vast list of references related to this topic. Schunker and Braun [7] have examined the power distribution for velocity observations from MDI and found that the excess high-frequency power corresponds to regions with the magnetic-field inclination (as deduced from potential-field source-surface extrapolation) in the 40–60 degree range. Jain and Haber [4] showed that halos seen in Dopplergrams from the MDI/SOHO satellite tend to be prominent in intermediate magnetic-field strengths of 50–250 G. The origin of these high-frequency halos remains unclear.

Jacoutot et al. [10] simulated the enhanced acoustic emission in active regions by investigating how magnetic field affected convective motion and excitation of solar oscillations. They concluded that localized enhancements seen in acoustic halos are caused by high-frequency turbulent convective motions in the presence of moderate magnetic field. Hanasoge [11] argued that the enhanced acoustic powers were a result of scattering by magnetic regions' low-$l$ to high-$l$ modes. It has been suggested [2] that acoustic emission from the solar granulation should be relatively localized and episodic, emanating largely as relatively discrete wavepackets emitted from convective plumes falling into the solar interior from near-surface layers at which granular convection takes place (see also [12]).

A computational seismic holography technique has been also used to search for additional emission of acoustic waves relative to the quiet Sun. The seismic source power computations consist in regressing the acoustic field a single skip, from the observed surface disturbance back to the surface point at which it is supposed to have originated. The resulting 6 mHz emission power of outward propagating waves reveal a seismic source excess surrounding the active region called the "seismic glory" ([6, 13]). The seismic glories contain some of the most intense seismic emitters, emitting as much as 150–250% more acoustic power than the quiet Sun.

Brown et al. [2] suggested that seismic emission originates from enhanced episodic source activity, as a characteristic of wave generation by turbulence. Donea et al. [6] found no evidence of this in temporal and spatial distributions of seismic-source episodes, either in acoustic-power halos or acoustic-emission halos. Recently Lindsey and Donea (2013, this proceedings) have applied seismic holography to high-frequency p-modes in the quiet Sun to study the statistical relationship between seismic sources and the structure of the solar granulation. For details on the general technique of seismic holography, we refer the reader to [13].

Khomenko and Collados [9] proposed that seismic halos can be caused by additional energy injected by high-frequency fast mode waves refracted in the higher atmosphere due to the rapid increase of the Alfvén speed. Acoustic waves trapped under magnetic canopy have been also suggested as a source of seismic power [14].

In this paper, we explore spatial properties of acoustic halos and seismic glories, as a function of wave frequency, inclination and strength of magnetic field, in order to understand the origin of seismic excess emission surrounding large active regions.

We use 24-hours of Doppler velocity and vector magnetogram observations from the Helioseismic and Magnetic Imager (HMI) on the Solar Dynamics Observatory (SDO) of the active region AR 11429 which generated a X5.4 solar flare on March 7, 2012.

## 2. Comparison of helioseismic observables

Tracked Doppler data cubes from HMI were Fourier transformed in time and acoustic power maps calculated in the frequency ranges 2.5–4.5, 5.5–6.5, 6.5–7.5, and 7.5–8.5 mHz. Seismic holography is also applied to the Fourier transforms to derive source power maps (left panels in Figure 1). For details on the general technique of seismic holography, we refer to [13]. For details on the application of the holography technique to episodic phenomena, we refer to [6]. The panels of Figure 1 show seismic source power maps (left) and acoustic power maps (right) in the frequency bands centered at 9 (top row), 8, 6, and 3.5 mHz (bottom row) for March 7, 2012. Both seismic-source power (left column) and local seismic power (right column) are seen to be suppressed in the active region at all frequencies; this is coincident with the location of moderate-to-strong magnetic fields, and includes plages. A high-frequency enhancement of power in both local seismic amplitude and local seismic emission surrounds the active region. The regions of enhanced local acoustic amplitude have been called "acoustic power halos". The regions of enhanced seismic emission have been called "acoustic glories". The seismic emission maps have a spatial resolution of 2.5 Mm, a diffraction limit attained by applying the diagnostic with an annular pupil (see Fig 4, p. 269 of [13]) whose radial range was 7–28 Mm. We found that the 6 mHz acoustic halo and the 6 mHz seismic-source glory streched over a region approximately 80–100 Mm thick.

The seismic source density is $|H_+|^2$. The mean of this for the quiet Sun is signified by $|H_{+,0}|^2$. Figure 2 shows the l-o-s magnetic field map with yellow contours indicating where normalized 6 mHz seismic-source power reverts to its quiet-Sun value, $|H_+|^2/|H_{+,0}|^2 = 1$. Green contours indicate the most seismic areas of the glories (exceeding 150% of $H_{+,0}$). Similarly, these zones also show the greatest enhancement in local acoustic power (50% above the quiet-Sun power). Zone 2 was found to be the most acoustically active at high frequencies. Figure 3 shows a slice of the acoustic emission maps centred on zone 2. The multifrequency acoustic and seismic emissions from zones 1 and 3 are shown in Figure 4. Table 1 shows a summary of the total acoustic power averaged over the defined areas, along with estimates of the mean magnetic field strength in the areas, and mean inclination angle of **B**.

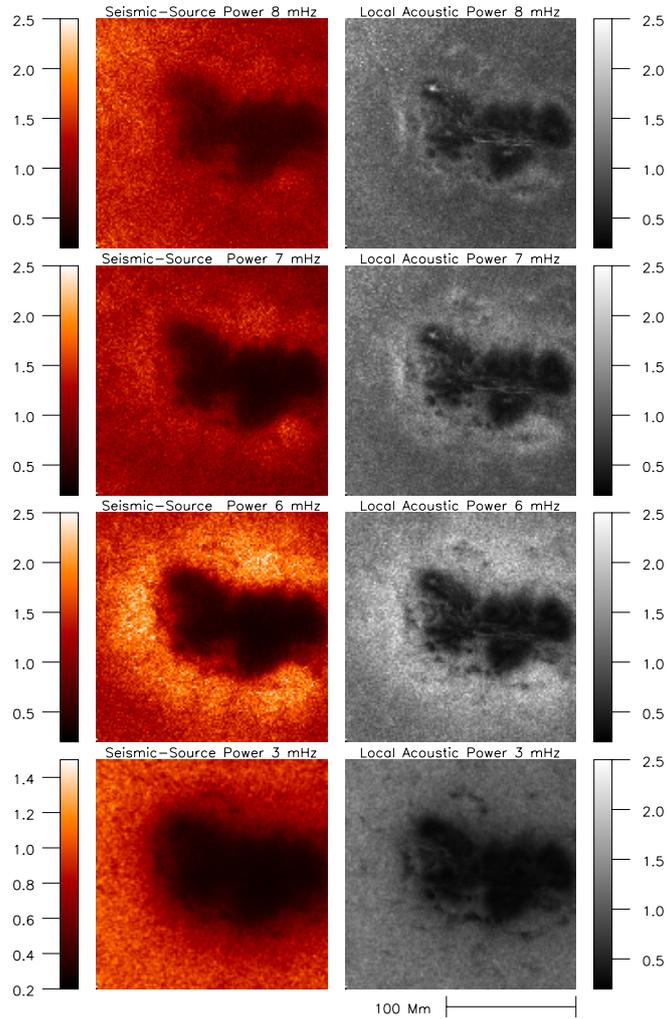

**Figure 1.** Multi-frequency oscillation power maps (as labelled) from seismic source emission (left column) and local Doppler velocity measurements (acoustic-power maps; right column). Maps are normalized to one in the quiet Sun.

We have identified a region of 6 mHz seismic emission deficit power ("quiet seismic moat") just beyond the white-light boundary of the active region (dashed yellow contours in Figure 2), surrounding even the most ragged areas of its boundary. It has a width of ≈ 20–30 Mm, and a highly inclined magnetic field (averaged magnetic field strength $B = 230$ G). The local seismic power in the moat ranges from 0.5–1 (in normalized units). The moat contains most of the plages surrounding the active region, clearly showing patchy features in acoustic/seismic emission due to the magnetic field acoustic attenuation. The magnetic canopy of the active region [16] may also have a role in changing the properties of solar convection in the moat, reducing the seismic emission. The seismic moat is surrounded by the seismic glory seen in all high frequency seismic power maps. We also found that the deficit in the seismic emission at the white-light boundary of the active region stays at a constant value of 0.5 (dashed yellow contours) of the quiet-Sun power everywhere around the active region.

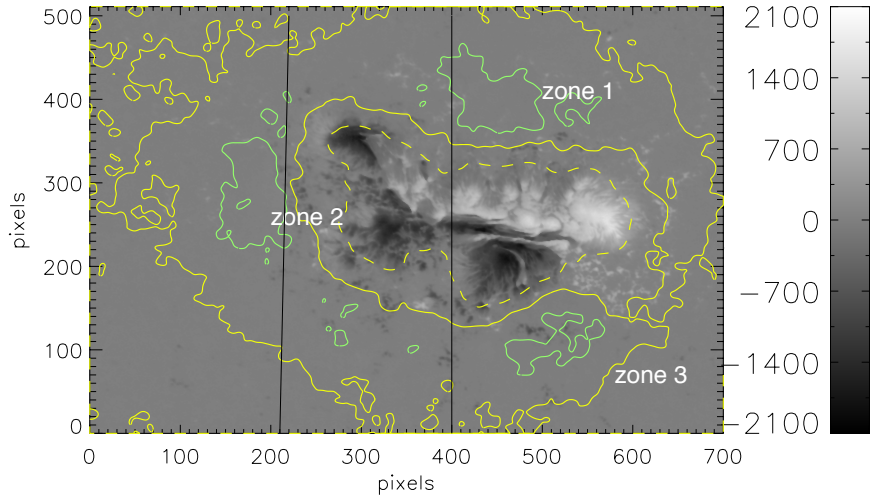

**Figure 2.** Contours of the 6 mHz seismic emission overplotted on the HMI l-o-s magnetic field strength map. Yellow contours show the seismic glory boundary, at the values of the quiet sun $|H_+|^2/|H_{+,0}|^2 = 1$. Dashed yellow curves show the white-light boundary of the active region. Green contours ($1.5 \times |H_{+,0}|^2$) show the most intense 6 mHz seismic areas, labelled as zone 1,2 and 3. Vertical lines define the submaps of the next figure.

Cospatial maps of 6 mHz local acoustic power, seismic source emission, continuum intensity, magnetic field strength from a vector magnetogram and field inclination angles $\gamma$ are shown in Figure 5. These images are centred upon zone 2 in Figure 2, the region of most intense seismic activity. We noticed an increased acoustic and seismic power in regions of approximately horizontal magnetic field whose strengths are between 50–200 G.

We are particularly interested in spatial correlations between the powerful discrete seismic emitters seen in glories and in the halos. Seismic holography is specifically designed for optimum local discrimination of acoustic sources, both temporally and spatially. Contours in Figure 5 show kernels in the 6 mHz local acoustic power (white) and 6 mHz seismic emission (black), within which the respective field value exceeds 180% of the quiet-Sun mean. Although in some areas we have identified a very good spatial correlation between the excess-local-seismic-power kernels and the excess-seismic-emission kernels, more than 60% of the respective kernels do not overlap. The highly emissive seismic sources have an average size of 2.0 Mm×3.5 Mm. To study the episodic nature of seismic emission in the glories we are preparing another paper, Hanson and Donea (2013).

We noticed a preference for the 7, 8 and 9 mHz enhanced acoustic power to shrink to a much narrower area outlining the active region and magnetic-field concentrations such as plages. Howe et al. [15] found similar behaviour for a different active region. We find that the corresponding regions of enhanced seismic emission at these frequencies become rather broad and diffuse by comparison (Figure 3), fading beyond 9 mHz. This finding is reflected also in Figure 6, where histogram plots emphasizing the difference in density distribution of the acoustic power halo and the seismic power glory at various frequencies are shown. The seismic emissivity dominates over the local acoustic oscillations at 6 mHz, reaching power values of 2.4×quiet-Sun.

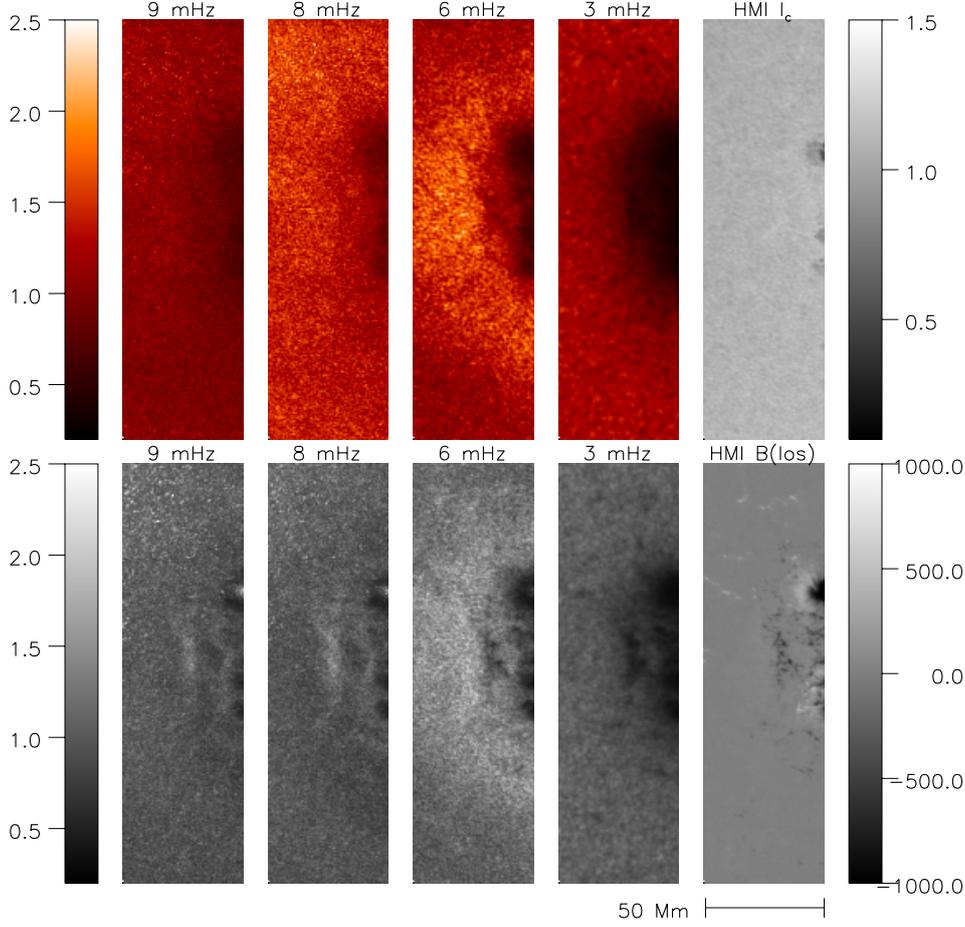

**Figure 3.** Seismic emission and acoustic local power maps at various frequencies in the most acoustically active part of the active region (zone 2). For reference, a normalized-to-the-quiet-Sun map of the intensity continuum and of the l-o-s magnetic field are included too.

The correlation of the 6-mHz local seismic power and seismic-source power with the strength and inclination of the magnetic field is expressed in zones 1, 2, and 3 by scatter plots in Figure 7. The 6 mHz seismic source power increases up to 240% of the quiet-Sun power at 6 mHz and is strongest at horizontal intermediate magnetic-field strength (120G).

We also computed the linear Pearson correlation coefficient between the 6 mHz acoustic power and 6 mHz seismic power in various frequency bands, for different zones of the halo. We find that there is a very good $R_{6\text{mHz}} = 0.979$ correlation between the 6 mHz acoustic power and seismic emission for the whole seismic halo.

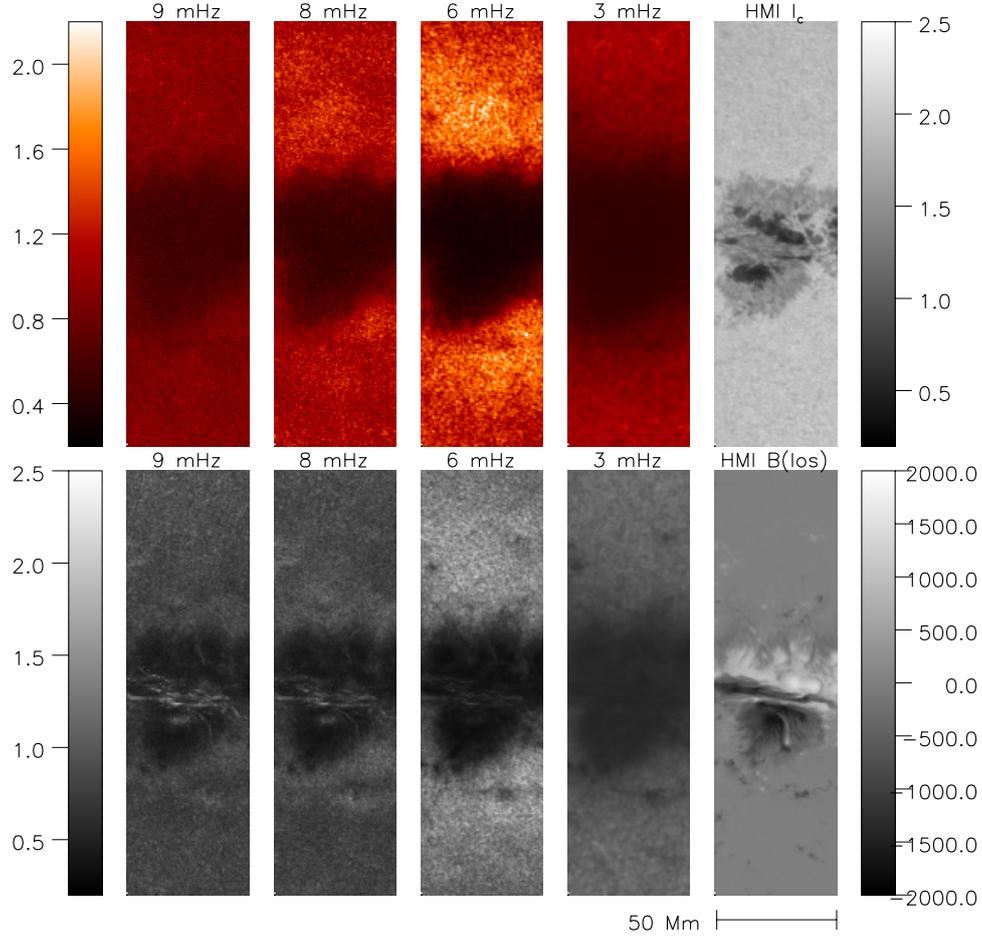

**Figure 4.** Same as the Figure 3, but for zone 1 (top) and zone 3 (below) the active region).

**Table 1.** Mean values of the Normalized Acoustic and Seismic Power of the most important areas of in the seismic halo. An average value for the total magnetic field strength and the inclination angle is also estimated.

| *Region* | *Zone 1* | *Zone 2* | *Zone 3* | *Halo* | *QuietSun* | *Moat* |
|---|---|---|---|---|---|---|
| Egression Power 3mHz | 0.72 | 0.96 | 0.80 | 0.88 | 1 | 0.44 |
| Acoustic Power 3mHz | 0.79 | 1.00 | 0.79 | 0.88 | 1 | 0.55 |
| Egression Power 6mHz | 1.91 | 9.91 | 1.73 | 1.55 | 1 | 1.27 |
| Acoustic Power 6mHz | 1.73 | 1.80 | 1.53 | 1.53 | 1 | 1.33 |
| Egression Power 8mHz | 1.14 | 1.34 | 1.00 | 1.07 | 1 | 0.83 |
| Acoustic Power 8mHz | 1.10 | 1.28 | 0.97 | 1.10 | 1 | 0.90 |
| HMI $B_{total}$ (G) | 148 | 139 | 141 | 147 | 116 | 230 |
| $\gamma$ (°) | 92 | 85 | 98 | 90 | 89 | 80 |

## 3. Final remarks

The purpose of this study is to examine the morphology of the enhanced seismic activity at high frequencies surrounding the region AR11429. We have found a very good spatial correlation between the 6 mHz glories and the 6 mHz power in the local oscillations of the surface. However, the spatial correlation weakens when one compares the compact kernels of the most intense local oscillation and seismic emission (180% of the respective quiet Sun values). The latter are dependent of the diffraction resolution of the seismic-emission maps, as rendered by helioseismic holography. The 7–28 Mm pupil collects Doppler signal from magnetic areas of the active region, where the acoustic oscillations are attenuated. This "showerglass effect" [16, 17] will have an impact on the assessment of acoustic emissivity and its relationship to local acoustic amplitude. Efforts to correct the "shower glass" effect could answer some of these issues. Our conclusions are:

- As expected, the local oscillation power and any seismic source activity are suppressed in the active region at all frequencies.

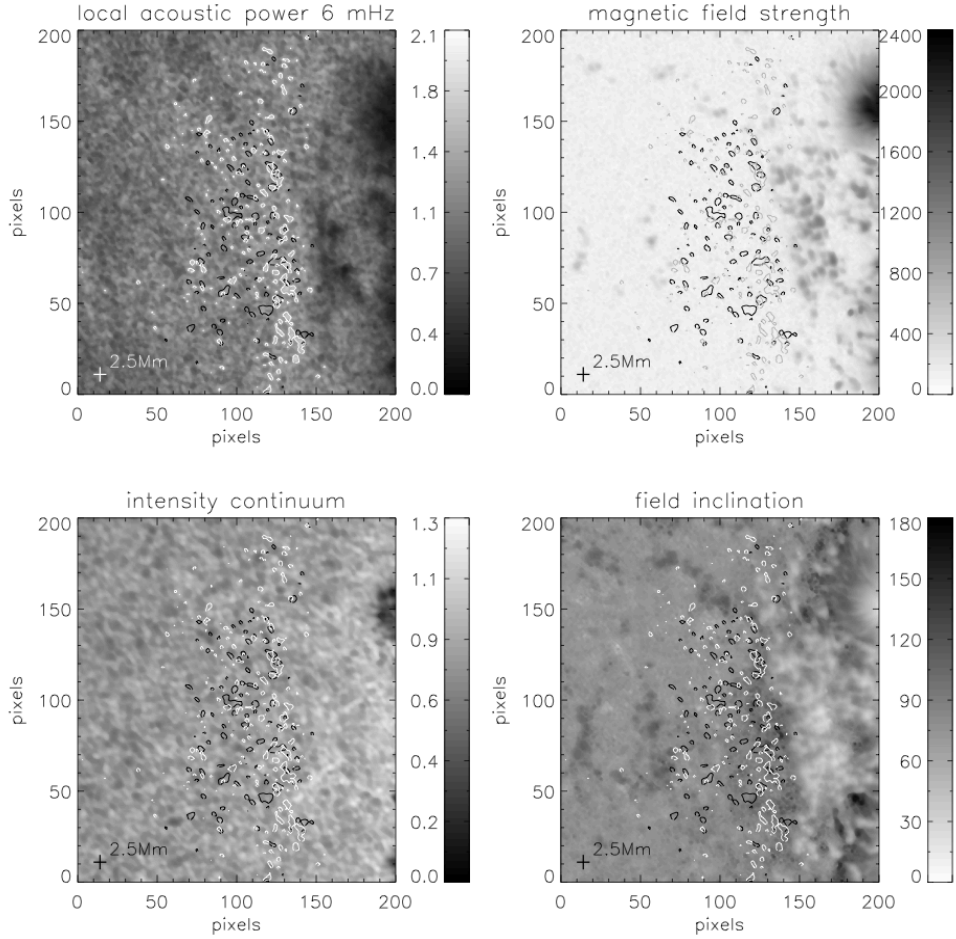

**Figure 5.** Co-spatial maps of zone 2 showing: a normalized 6 mHz local acoustic power, HMI/SDO vector magnetic field strength, normalized intensity continuum and field inclination $\gamma$. A weak spatial correlation exists between the acoustic power averaging roughly 180% in excess of the mean quiet Sun local acoustic power (white contours) and the seismic power at 180% in excess of the quiet Sun seismic power (black contours). Most of high power seismic emitters thrive in the areas free of plages, surrounding the active region. The length of each side of the maps is $\approx 70$ Mm (1 pixel = 0.3475 Mm). The cross shows the spatial resolution of 2.5 Mm of the 6mHz seismic source power maps.

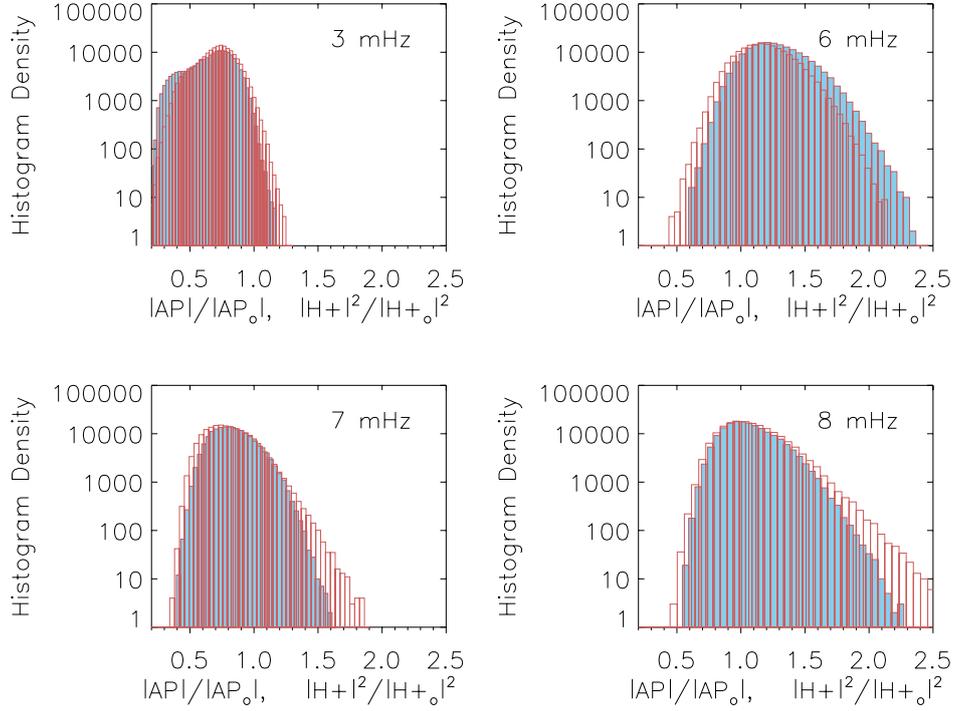

**Figure 6.** Histogram plots of the normalized acoustic power $|AP|/AP_o|$ (red) and seismic power $|H_+|^2/|H_{+,0}|^2$ (blue). At 6 mHz the seismic source emission dominates over the local acoustic oscillations.

- These regions of suppression are surrounded by halos of enhanced local-seismic-amplitude/power as well as enhanced seismic emission.
- The seismic-power halo contracts and weakens with increasing frequency, becoming indistinct by about 10 mHz. The seismic-emission halo (i.e., the "glory"), likewise weakens with increasing frequency, but, rather than shrinking, becomes more diffuse at higher frequencies.
- Seismic emission from areas surrounding the active region reaches up to 240% of the quiet Sun power in compact kernels.
- The patchiness of the seismic source density distribution in acoustic glories (and halos) appears to be related to plages.
- The seismic source power in the glory dominates over the local surface oscillations at frequencies of 6 mHz by a factor of 25%. This dominance appears to change with frequency.
- The white-light boundary of the active region (sometimes where the penumbra of large sunspots reaches into the quiet Sun) sustains a local seismic power and seismic emissivity of 0.5 of the respective mean-quiet-Sun power and emissivity.
- We recognise a "moat" of 6 mHz seismic emission deficit extending 20–30 Mm wide outside of the active region boundary.

There appears to be a strong relationship between the enhanced acoustic power seen at high frequencies and the enhanced seismic emissivity. However, differences between the distributions of the two phenomena suggest that this relationship may be somewhat oblique. If the sensitivity of the photosphere to upcoming seismic radiation were the same in seismic halos as in the quiet Sun, we might expect a closer relationship, especially at high frequencies. If, for example, local emitters submerged just beneath the photosphere emitted the same power upward, into the local overlying photosphere, as downward (to arrive into the distant pupil of a holographic diagnostic), then we could expect regions of enhanced seismic power (directly above acoustic emitters) to coincide closely with regions of enhanced acoustic emission (into distant pupils). We understand that this would be the case for simple dipole or quadrupole emitters. However, it is possible to contrive coherently related combinations of dipole and quadrupole emitters that violate this relationship.

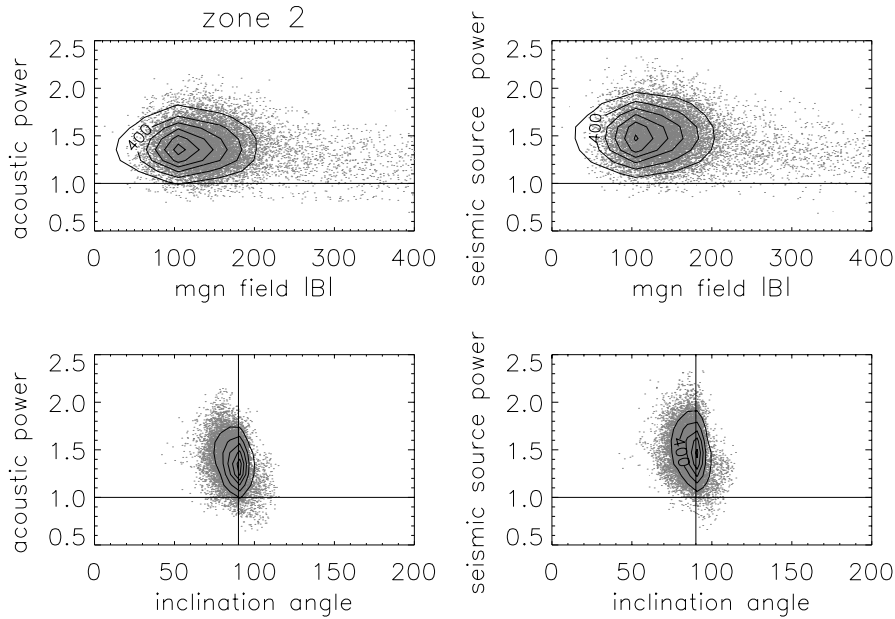

**Figure 7.** Scatter plot of normalized 6 mHz acoustic and seismic source power in zone 2, versus the magnetic field strength and magnetic field $\gamma$ inclination angles. Vertical lines indicate $\gamma = 90°$. Power maps are normalized to one in the quiet-Sun.

To complicate the issue, it is also possible that regions of quasi-horizontal magnetic fields with strength 50–200 Gauss have a different sensitivity to seismic noise arriving in their environments. Simulations of the effects of inclined magnetic field on helioseismic signatures could shed some light on this question.

We thank the referee for very useful comments and suggestions.

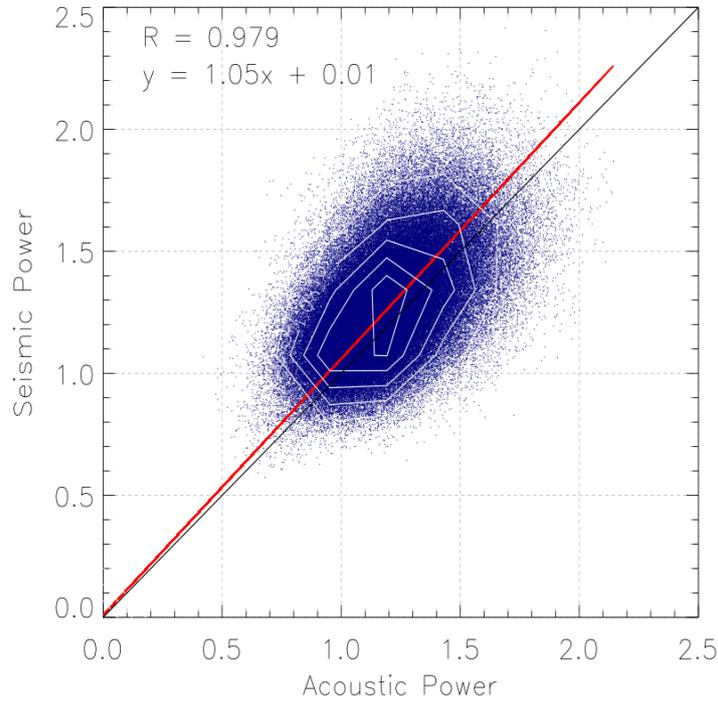

**Figure 8.** Scatter plot of normalized 6 mHz acoustic and seismic source power in the entire seismic glory. Contours show the density distribution of points, with a higher density at position (1.3,1.3). The Pearson correlation coefficient and the equation of the fitted line (thin red line) is displayed on the scatter plot. The black diagonally line is drawn for the purpose of seeing the slight deviation of the scatter plot from a perfect linear correlation.